\documentclass[twocolumn, amsfonts, aps, pra]{revtex4}

\usepackage{epsfig}

\begin{document}

\title{Unified criterion 
for security of secret sharing in terms of violation of Bell inequalities}

 \author{Aditi Sen(De), Ujjwal Sen, and Marek \.Zukowski}
 \affiliation{Instytut Fizyki Teoretycznej i Astrofizyki, Uniwersytet
 Gda\'nski, PL-80-952 Gda\'nsk, Poland}

\begin{abstract}
In secret sharing protocols, a secret  is to be distributed among several partners
so that leaving out any  number of them, the rest do not have the complete information.
Strong multiqubit correlations in the state by which secret sharing is carried out,
had been proposed as a criterion for security of such protocols
against individual attacks by an eavesdropper. 
However we show that states with \emph{weak} multiqubit correlations can
 also be used for secure secret sharing.
 That our state has \emph{weak} multiqubit correlations, is shown
from the perspective of violation of local realism, and also by showing that its higher 
order correlations are described by lower ones. 
We then present a unified criterion for security 
of secret sharing in terms of violation of local realism, which works when the secret 
sharing state is  the Greenberger-Horne-Zeilinger state 
(with \emph{strong} multiqubit correlations), as well as states of a different class
 (with \emph{weak}  multiqubit correlations). 
\end{abstract}

\maketitle

\newcommand{\tr}{{\rm tr}}

\section{Introduction}

Violation of Bell inequalities 
seem to be a signature of what has been called ``useful entanglement'' \cite{GS1, ASW}.
It was shown in Ref. \cite{GS1} that violation of local realism can be seen as a 
criterion 
for security of secret sharing protocols \cite{Zukowski, Hillery}. 
It was argued that a strong violation of multi-qubit Bell inequalities \cite{WWWZB, ZB}
by the state that acts as the vehicle in secret sharing can be a criterion for security of 
the secret sharing. 
Violation of local realism was also 
shown to be connected with distillability
\cite{ASW} and super-classical communication complexity \cite{complexity}.

In a secret sharing protocol,  the holder of a secret (call her Alice)
wants to distribute her secret among \(N-1\) separated parties (call them Bobs, \(B_1,
B_2, \ldots, B_{N-1}\)) such that leaving out any (non-zero) number of the Bobs,  
the other Bobs would have 
 no information about Alice's secret.
Such protocols were shown to be possible in Refs. \cite{Zukowski, Hillery},
if Alice and the \(N-1\) Bobs share a large 
number of 
certain entangled states \cite{quantumsecret,quantumsecretreferences}. 

Such a protocol could suffer 
from the onslaught of a possible eavesdropper Evan. Evan 
could spy (quantum mechanically) on the channels that carry the states from 
Alice to the Bobs and obtain information about Alice's secret.
In Ref. \cite{GS1}, it was shown that if the secret sharing is carried out by 
using Greenberger-Horne-Zeilinger (GHZ) states,
the secret of Alice is 
secure from eavesdropping as long as the state between Alice and Evan (after eavesdropping)
does not violate any Bell inequality while the state between Alice and the Bobs
(after eavesdropping) violates the \(N\)-qubit  Bell inequalities \cite{WWWZB, ZB} 
very strongly. 

In this paper, we show that security of secret sharing in the multi-qubit scenario
can be possible even when
the state between Alice and the Bobs 
of the secret sharing protocol contains weak multi-qubit correlations.
We argue that the criterion for security of secret sharing is  
of a different type. 
\begin{enumerate}
\item[(i)] The state shared between Alice and the Bobs (after eavesdropping), 
if suitably projected into certain states by all but one Bobs, the remaining Bob
would share a state with Alice, which violates a two-qubit Bell 
inequality.
\item[(ii)] At the same time, the state between Alice and Evan must satisfy such inequalities.  
\end{enumerate}
We show that both for 
the GHZ state, which contains
\emph{strong} multi-qubit correlations (in the sense of 
strong violation of multi-qubit Bell inequalities),  as well as for
another state (we call it the \(G\) state), which in several ways (as indicated below) contains \emph{weak} 
multi-qubit correlations, 
 the security of secret sharing is exactly in the same range in which 
both the above conditions, (i) and (ii),
are met. 
This criterion  can therefore
be seen as a \emph{unified} criterion for security of secret sharing.
We subsequently show, that the \(N\)-qubit \(G\) 
state cannot have strong 
\(N\)-qubit correlations. This is in the sense that 
\begin{enumerate}
\item[(a)] the 
\(N\)-qubit correlation functions of the \(G\)-state
only weakly violate the multi-qubit  Bell inequalities \cite{WWWZB, ZB}
as compared to the  GHZ state.
\item[(b)] The \(N\)-qubit correlations of  this state 
are determined by lower-order correlations of the state in contrast to 
the GHZ state \cite{LPW}. 
\end{enumerate}

\section{New states for secret sharing}

The state 
\begin{equation}
\label{G}
\left|G_N\right\rangle = \frac{1}{\sqrt{2}}(\left|W_N\right\rangle + \left|\overline{W}_N\right\rangle)
\end{equation}
can be used 
for secret sharing,
where
\[
\begin{array}{lcl}
\left|W_N\right\rangle = \frac{1}{\sqrt{N}}\sum \left|10^{\otimes N-1}\right\rangle, \\
\left|\overline{W}_N\right\rangle = \frac{1}{\sqrt{N}}\sum \left|01^{\otimes N-1}\right\rangle,
\end{array}
\]
with \(\left|0\right\rangle\) and \(\left|1\right\rangle\) being
the eigenvectors of \(\sigma_z\),
and  e.g., \(\sum \left|10^{\otimes N-1}\right\rangle\) denotes
the unnormalised superposition of all \(N\) arrays
of \((N-1)\) \(\left|0\right\rangle\)s and a single \(\left|1\right\rangle\). 
Note that 
\[\left|G_{4}\right\rangle = \frac{1}{\sqrt{2}}(\left|+x \right\rangle^{\otimes 4} - 
\left|-x \right\rangle^{\otimes 4})\] where \[\left| \pm x \right\rangle = 
\frac{1}{\sqrt{2}}(\left|0\right\rangle \pm \left|1\right\rangle).\]
However, for more qubits, this family is different than the GHZ family, as we would show later.

The property of the \(G\) state that would help us to use it for secret sharing 
is the 
following:
\begin{equation}
\label{main}
\left\langle \sigma_x^{\otimes N}\right\rangle_{G_N}  =  1, \quad
\left\langle \sigma_y^{\otimes N}\right\rangle_{G_N}  =  (-1)^{M+1},
\end{equation}
\emph{whenever} \(N = 2M\).
The first equation follows from the fact that 
\[\sigma_x^{\otimes N} \left|W_N\right\rangle = \left|\overline{W}_N\right\rangle, \quad 
\sigma_x^{\otimes N} \left|\overline{W}_N\right\rangle = \left|W_N\right\rangle.  
\]
To derive the second equation in eq. (\ref{main}), note that 
\[\sigma_y^{\otimes N} \left|G_N\right\rangle = i^N\frac{1}{\sqrt{2}}(-\left|\overline{W}_N\right\rangle
 + (-1)^{N-1} \left|W_N\right\rangle).
\]
The state on the right-hand-side  is orthogonal to \(\left|G_N\right\rangle\)
 for odd \(N\). But for even \(N (=2M)\), the right-hand-side is 
\(-i^{2M} \left|G_N \right\rangle\).

Suppose that Alice (A) has some information which she wishes to  secretly distribute 
among \(2M-1\) Bobs (\(B_1, B_2, \ldots , B_{2M-1}\)). But she wishes to 
distribute it in such a way that 
\emph{all} the Bobs must cooperate to obtain the message. Leaving out any nonzero 
number of Bobs, the remaining 
Bobs would not be able to gather the complete  information about Alice's bit. 
This is equivalent to the situation when Alice  is able to 
secretly distribute a random sequence of binary digits (bits)
to the Bobs with the same property. 
Let us now show  that this is possible by using the \(G_{2M}\)-state,
shared between Alice and the \(2M-1\) Bobs.  Suppose that 
Alice and the Bobs share 
a large number of the \(G_{2M}\)-state and they 
randomly choose between the \(\sigma_x\) and \(\sigma_y\) observables to make a 
measurement on their respective parts of the shared states \cite{Ekert}.  
Subsequently they publicly share  information about the \emph{bases} in which 
they have made their measurements. (They obviously 
do not share information about the \emph{results} of the measurements.) They keep 
only those results in which 
\emph{all} of them measured in the same basis, that is when either all of them measured 
in the 
\(\sigma_x\)-basis or all of them measured in the \(\sigma_y\)-basis. 
In any run of such a Bell type experiment,
if \emph{all} the parties happen to choose \(\sigma_x\) as their observable, then the \(2M\) 
results
at Alice and the \(2M -1\) Bobs are related by (see eq.(\ref{main}))
\[r_A r_{B_1}  \ldots r_{B_{2M-1}} =1.\]
If all the Bobs cooperate, then they know that \[r_A = r_{B_1}  \ldots r_{B_{2M-1}}.\]
If any nonzero number of Bobs refuse to cooperate or are left out, the remaining Bobs  
cannot gather the complete information about \(r_A\). 
This is because irrespective of what results that any \(2M-2\) Bobs happen to obtain
in their \(\sigma_x\) basis measurements, the state of remaining two qubits 
is such that the single-qubit density matrix at Alice is mixed.
Similarly, if \emph{all} the parties happen to choose \(\sigma_y\)
as their observable, then the 
result of Alice is related to those of the Bobs by
\[s_A = (-1)^{M+1}  s_{B_1}  \ldots s_{B_{2M-1}}.\] In this way, a sequence
of random bits is created at Alice and if there is no eavesdropping, the Bobs 
(if all of them 
cooperate) would be able to reproduce the same random sequence.

\section{Security}

We now consider the security of the distributed sequence of random bits against a 
possible eavesdropper, Evan.  We work with the assumption that 
Evan would only be able to make coherent but individual attacks. Thus  although Evan 
may attack coherently on all the qubits sent to the Bobs (from Alice) in a single run 
of the experiment, he is not able to perform joint
operations on qubits from different runs. As has been stressed in a recent
review \cite{rmp}, at present, even individual attacks by an eavesdropper would be very challenging in reality.

The states that are shared by   
the Bobs after the measurements by Alice are as follows:
\begin{equation}
\label{Z}
\begin{array}{rcl}
\left|\pm x\right\rangle & : & \frac{1}{\sqrt{2}}(\left|\xi\right\rangle \pm \left|\overline{\xi}\right\rangle) \\
\left|\pm y\right\rangle & : & \frac{1}{\sqrt{2}}(\left|\xi\right\rangle \pm i \left|\overline{\xi}\right\rangle), 
\end{array}
\end{equation}
the left hand column in eq.(\ref{Z})
being the outcome obtained at Alice, while the right hand side 
gives the corresponding state that is shared by the Bobs.
Here 
\(\left|\pm x \right\rangle\) and \(\left|\pm y \right\rangle\) are the eigenvectors
of \(\sigma_x\) and \(\sigma_y\), and
\begin{equation}
\label{+Z}
\begin{array}{rcl}
\left|\xi\right\rangle &=& \frac{1}{\sqrt{2M}}(\sum \left|1 0^{\otimes 2M-2}\right\rangle 
                                                                            + \left|1^{\otimes 2M -1}\right\rangle)\\
\left|\overline{\xi}\right\rangle &= &\frac{1}{\sqrt{2M}}(\sum \left|0 1^{\otimes 2M-2}\right\rangle 
																										        + \left|0^{\otimes 2M -1}\right\rangle).
\end{array}
\end{equation}

Note that the protocol for secret sharing 
is exactly equivalent to the BB84 key distribution protocol \cite{BB84}, when
we consider the problem of eavesdropping and the eavesdropping is coherent. 
Consequently the optimal coherent individual attack that Evan (\(E\)) can perform, is given by the 
  unitary transformation \cite{Fuchs}
\begin{equation}
\begin{array}{rcl}
\label{unitary}
U_{BE} \left|\xi\right\rangle\left|0\right\rangle & = &     \left|\xi\right\rangle\left|0\right\rangle\\ 
U_{BE} \left|\overline{\xi}\right\rangle\left|0\right\rangle & = & \cos \phi \left|\overline{\xi}\right\rangle\left|0\right\rangle
	                                                     +  \sin \phi \left|\xi\right\rangle\left|1\right\rangle,
\end{array}
\end{equation} 
 where \(B\) denotes \(B_1B_2 \ldots B_{2M-1}\) and \(\phi \in [0,\pi/2]\). 
After Evan implements his attack, i.e., 
after he  applies the unitary operation \(U_{BE}\) given by eq.(\ref{unitary}), the state  \(G_N\) of
Alice and the Bobs coupled with the probe \(\left|0\right\rangle\) of Evan transforms as
\begin{equation}
\begin{array}{rcl}
\label{three}
\left|G_N\right\rangle_{AB}\left|0\right\rangle_E \rightarrow 
\left|\psi\right\rangle_{ABE} & = &
                         \frac{1}{\sqrt{2}}(\left|0\right\rangle_A \left|\xi\right\rangle_B\left|0\right\rangle_E \\
                        + \cos \phi \left|1\right\rangle_A \left|\overline{\xi}\right\rangle_B\left|0\right\rangle_E                              
                         & + & \sin \phi \left|1\right\rangle_A \left|\xi\right\rangle_B\left|1\right\rangle_E).
\end{array}
\end{equation}

The condition that is needed for security is \cite{security}
\begin{equation}
\label{security}
I(A:B) > I(A:E),
\end{equation}
where again \(B\) denotes 
the aggregate of 
all the Bobs. This security implies that Alice and the Bobs can run a one-way protocol (privacy 
amplification) if and only if the condition (\ref{security}) is satisfied. 
Here the mutual information \(I(X:Y)\) is defined as 
\[I(X:Y) = H(X) - H(X|Y),\] where \[H(\{p_i\}) = - \sum_ip_i\log_2p_i\] is the Shannon entropy
of a probability distribution \(\{p_i\}\).

We would now calculate the mutual information (after Evan's attack), when
Alice and the Bobs measure either all of them in the \(\sigma_x\) basis or all of them in 
the \(\sigma_y\) basis. This happens with equal probability.
Therefore in our case, \[I(A:B) = 1 - \frac{1}{2}(H_x(A|B) + H_y(A|B)).\]
(As Alice chooses her measurements randomly, \(H(A) =1\).)
The state shared between Alice and the \(2M -1\) Bobs (after Evan's attack) is (see eq.(\ref{three}))
\begin{equation}
\label{rhoab}
\rho_{AB} 
 = \frac{1+\cos^2\phi}{2} \left|\alpha \right\rangle \left\langle \alpha \right|
 +   \frac{\sin^{2} \phi}{2} \left|1\right\rangle\left| \xi \right\rangle 
                                             \left\langle 1 \right| \left\langle \xi \right|,
\end{equation}
where 
\[\left|\alpha \right\rangle = \frac{\left|0 \right\rangle \left|\xi\right\rangle 
         + 
           \cos\phi \left|1 \right\rangle \left|\overline{\xi} \right\rangle}{\sqrt{1 + \cos^2\phi}}.\]
Let us first evaluate
\begin{equation}
\label{hx}
\begin{array}{rcl}
H_x(A|B) &\equiv& p_x(B=1) H_x(A|B=1) \\
&+& p_x(B=-1) H_x(A|B=-1),
\end{array} 
\end{equation}
where for example, \(p_x(B = 1)\) is the probability of the event ``B = 1'', when all the Bobs
measure in the \(\sigma_x\) basis.

Tracing out  \(A\) from \(\rho_{AB}\), the state \(\rho_B\) shared between the \(2M-1\) 
Bobs is (see eq.(\ref{rhoab}))
\[\frac{1 + \sin^2 \phi}{2} \left|\xi\right\rangle \left\langle \xi \right|
                                    + \frac{\cos^2 \phi}{2} \left|\overline{\xi}\right\rangle 
               																					\left\langle \overline{\xi} \right|.\]
 Consequently,
\[p_x(B= \pm 1) = \frac{1}{2}\]
and from eq.(\ref{rhoab}), 
\[
p_x(A=1, B= \pm  1) = p_x(A=-1, B= \mp 1) =\frac{1 \pm \cos \phi}{4} 
\]
Therefore 
\[
H_x(A|B=1)
 = H(\frac{1 + \cos \phi}{2}),\] 
where \(H(p) = -p\log_2p -(1-p)\log_2(1-p)\) is the binary entropy function.
By symmetry, the expression for \(H_x( A|B=-1)\) is exactly the same.
So from eq.(\ref{hx}), one obtains 
\[
H_x(A|B) = H(\frac{1 + \cos \phi}{2})
\]
By symmetry, the expression for \(H_y(A|B)\) is exactly the same 
as that for \(H_x(A|B)\).
So finally,
\begin{equation}
\label{Iab}
I(A:B)=  1 -  H(\frac{1 + \cos \phi}{2}).
\end{equation}
\(I(A:E)\) is obtained by replacing \(\phi\) by \(\pi/2 - \phi\) in 
\(I(A:B)\).
With these  expressions, one can see 
that the security condition (\ref{security}) holds if and only if
\(\phi < \pi/4\).

Consider now the state \(\rho_{AB}\) (given by eq.(\ref{rhoab})), obtained by tracing out the eavesdropper \(E\),
after the eavesdropping. If any \(2M-2\) of the \(2M -1\) Bobs perform
measurements in the \(\sigma_z\) basis and obtains either \(\left|0\right\rangle^{\otimes 2M-2}\)
or \(\left|1\right\rangle^{\otimes 2M-2}\) (cf. \cite{Popescu, ZHG}),
 then the remaining Bob (say \(B_k\)) shares with Alice the 
(two-qubit) state
\[\rho_{AB_{k}} = \frac{1 + \cos^2 \phi}{2} \left|\beta\right\rangle \left\langle \beta \right|
+ \frac{\sin^2\phi}{2} \left|11 \right\rangle \left\langle 11 \right|,\] 
where 
\[ \left| \beta \right\rangle = \frac{\left|01\right\rangle + \cos\phi \left|10 \right\rangle}
{\sqrt{1 + \cos^2 \phi}}.
\]
The two-qubit state \(\rho_{AB_k}\) violates local realism if and only if \(\phi < \pi/4\) \cite{HHH}. 
Note that there are  no 
sequential measurements involved. The parties follow the ordinary Bell-type experiment 
and classical communication is needed only to share this data. We consider 
violation of local realism exhibited by a subset of this data. Measurements at the parties do not depend 
on results of measurements at the other parties.

On the other hand, the state obtained after we trace out the Bobs
(from the state \(\left|\psi\right\rangle_{ABE}\), given by eq. (\ref{three})), is 
\[\rho_{AE} = \frac{1 + \sin^2 \phi}{2} \left|\gamma\right\rangle \left\langle \gamma \right|
+ \frac{\cos^2\phi}{2} \left|10 \right\rangle \left\langle 10\right|,\] 
where 
\[\left|\gamma\right\rangle = \frac{\left|00\right\rangle + \sin\phi \left|11 \right\rangle}
{\sqrt{1 + \sin^2 \phi}}.
\]
The state \(\rho_{AE}\) violates local realism if and only if \(\phi > \pi/4\) \cite{HHH}.

Therefore for \(\phi \in [0, \pi/4)\), the state shared (after eavesdropping by Evan) 
between Alice and the Bobs violate local realism
in the sense described above, while the Alice-Evan state  does not violate local realism. 
And the secret sharing is secure in exactly the same range.

\section{Comparison with the GHZ state}

A related feature was obtained in \cite{GS1}, where 
 secret sharing was investigated by using the GHZ state \cite{robust}
\[\left|\mbox{GHZ}_N\right\rangle = \frac{1}{\sqrt{2}}(\left|0^{\otimes N}\right\rangle 
                                     + \left|1^{\otimes N}\right\rangle).\]
However in this case, 
it was shown that the secret sharing protocol is secure as long as after eavesdropping by Evan,
 strong \(N=2M\) qubits correlations are still present in the state
shared by Alice and Bobs.
If secret sharing is carried out by a shared GHZ state, as has been considered in  \cite{GS1},
the shared state (after optimal eavesdropping by Evan) between Alice, the 
\(2M -1\) Bobs and Evan is just the same
as displayed on the right hand side of eq.(\ref{three}) with the replacement 
\[
\begin{array}{lcl}
\left|\xi\right\rangle & \rightarrow & \left|0^{\otimes 2M - 1}\right\rangle \\
\left|\overline{\xi}\right\rangle & \rightarrow & \left|1^{\otimes 2M - 1}\right\rangle.
\end{array}
\]
Consider the state between Alice and any one of the  Bobs, after measurements in the \(\sigma_x\)-basis
at the other Bobs, and suppose that either \(\left|+ x\right\rangle\) clicks at all those \(2M -2\) Bobs,
or \(\left|- x\right\rangle\) clicks at all of them. As can be 
checked, this state again violates local 
realism  if and only if \(\phi < \pi/4\). The state shared by Alice and Evan does not violate local realism
in that range. This is  the range in which secret
    sharing is secure.

In  \cite{GS1}, it was numerically shown that 	the complete set of Bell inequalities 
in multi-qubit systems \cite{WWWZB, ZB} is violated by 
the state shared between Alice and the \(2M -1\) Bobs (after eavesdropping by Evan)
by a magnitude of more than \(2^{\frac{2M - 1}{2}}\)
if and only if \(\phi \in [0, \pi/4)\). In this  range, the Alice and Evan 
state does not violate any Bell inequality. It was therefore proposed 
that this form of strong violation of local realism for 
the \(2M\)-qubit correlations is a criterion for security in secret sharing.

However we have shown that the range in which security of secret sharing by a GHZ state is obtained, 
is also concurrent with violation of local realism in another form; that of satisfying items
(i) and (ii) in the Introduction. This feature is also shared by our \(G\) states. The proposed 
criterion is therefore a unified criterion  for security of secret sharing. 

\section{Strength of multiqubit correlations}

The GHZ states have strong 
multiqubit correlations. In contrast, the family  \(\left|G_{2M}\right\rangle\),  
which can also be considered as
the vehicle 
of the secret sharing,
is unlikely to 
produce a strong violation of 
local realism for \(2M\)-qubit correlations. On the contrary, we will show that it has a 
large amount of its entanglement
concentrated in correlations of 
lower number of parties. In the following, we drop the restriction that the number of parties 
is even. 

\subsection{Violation of local realism}

Consider the white noise admixed \(N\)-qubit state \(G_N\), 
\begin{equation}
\label{noisyGN}
\rho^{G_N} = p_N \left|G_N\right\rangle \left\langle G_N \right| + ( 1 - p_N) \rho^{N}_{noise}
\end{equation}
where \(\rho_{noise}^{N} = \frac{1}{2^N}I\) is the maximally mixed state of \(N\) qubits.
If \(N-2\) parties make  measurements in the
\(\sigma_z\) basis, and if either \(\left|0\right\rangle\) clicks at all the parties,
or \(\left|1\right\rangle\) clicks at all the parties,
the collapsed state  is given respectively by 
\[\rho^{G_2}\otimes(\otimes_{i=3}^{N} \left|0\right \rangle_{i i} \left\langle 0\right|)\] 
or 
\[\rho^{G_2}\otimes(\otimes_{i=3}^{N} \left|1\right \rangle_{i i} \left\langle 1\right|),\] 
where  
 \(\rho^{G_2}\) is the Werner state 
\[ \rho^{G_2} = p_{(N\rightarrow 2)} \left|G_2\right\rangle \left\langle G_2 \right|
                    + (1 - p_{(N\rightarrow 2)}) \rho^2_{noise}\]
with 
\[p_{(N\rightarrow 2)} = \frac{1}{1 + \frac{(1-p_N)N}{p_N 2^{N-2}}},\]
and 
\[
\left|G_2\right\rangle  = \frac{1}{\sqrt{2}}(\left|01\right\rangle + \left|10\right\rangle).
\]
The state \(\rho^{G_2}\) has no  local realistic description  for 
\(p_{(N\rightarrow 2)} > 1/\sqrt{2}\). This would imply that the state \(\rho^{G_N}\) cannot 
have a local realistic model for
\[ p_{N} > p_{G_N}^{crit} \equiv  \frac{N}{N + (\sqrt{2}-1)2^{N-2}}.\]

A GHZ state of \(N\) qubits 
maximally violates the multiqubit Bell  inequalities \cite{WWWZB, ZB} and the amount of 
violation is \(\sqrt{2^{N+1}}\). Consequently the noisy GHZ state 
\begin{equation}
\label{noisyGHZ}
\rho^{GHZ_N} = q_N P_{\left|GHZ_N\right\rangle} + (1 - q_N) \rho_{noise}^{N}
\end{equation}
violates local realism for \[q_N > q_{GHZ_N}^{crit} \equiv 1/\sqrt{2^{N-1}}.\]
But for \(N\geq 13\), \[p_{G_N}^{crit} > q_{GHZ_N}^{crit}.\] 
Thus 
for \(N \geq 13\), the nonclassicality of the correlations in \(G_{N}\) is more robust to ``white noise''
 admixture than 
the one for the  corresponding GHZ state. 
A similar method was used in Ref. \cite{ZHG} (cf. \cite{Popescu}) to show
that the nonclassical behavior of the \(W\) states is stronger than that 
of the GHZ state for sufficiently large number of parties. 
Note here that 
the robustness to white noise admixture of the nonclassicality of correlations in the \(G_N\) 
states obtained by the above method 
is weaker than the one that is obtained for the corresponding \(W\) state.

Therefore strong nonclassical properties of the \(G_N\) states emerge
 after local projection measurements in some parties. 
This strong nonclassicality for  lower number of parties seems to imply that the \(N\)-party 
correlations are weak. One can see this more directly
in the following considerations for the \(6\)-qubit \(G\) state, the 
lowest number of qubits in which  secret sharing is possible with a \(G\) state and which is 
not simply a GHZ state. Consider the correlation tensor
\(\hat{T}^{G_6}\)
of the state \(G_6\), the elements of which are given by 
\[T_{x_1 \ldots x_6} = \tr(P_{\left|G_6\right\rangle}\sigma_{x_1}^{(1)} 
\ldots \sigma_{x_6}^{(6)}), \quad (x_i = x, y, z).\]
Explicit calculation reveals the following structure of the tensor: 
\[
\begin{array}{lcl}
\hat{T}^{G_6}  =  \otimes_{i=1}^{6}\vec{x_i} + \otimes_{i=1}^{6}\vec{y_i} - \otimes_{i=1}^{6}\vec{z_i} \\
                +  \frac{1}{3}\{  \sum (\otimes_{i = 1}^{2}\vec{x_i}) \otimes (\otimes_{j = 3}^{6}\vec{z_j}) 
                      -     \sum (\otimes_{i = 1}^{4}\vec{x_i}) \otimes (\otimes_{j = 5}^{6}\vec{z_j}) \\
                      -     \sum (\otimes_{i = 1}^{4}\vec{y_i}) \otimes (\otimes_{j = 5}^{6}\vec{z_j}) 
                     +  \sum  (\otimes_{i = 1}^{2}\vec{y_i}) \otimes (\otimes_{j = 3}^{6}\vec{z_j}) \\
                     -   \sum (\otimes_{i = 1}^{2}\vec{x_i}) \otimes (\otimes_{j = 3}^{6}\vec{y_j}) 
                    -   \sum (\otimes_{i = 1}^{4}\vec{x_i}) \otimes (\otimes_{j = 5}^{6}\vec{y_j}) \\
       +   \sum (\otimes_{i = 1}^{2}\vec{x_i}) \otimes (\otimes_{j = 3}^{4}\vec{y_j}) 
\otimes (\otimes_{k = 5}^{6}\vec{z_k})\}, 
\end{array}
\]
where for example \(\sum \vec{x_1} \otimes \vec{x_2} \otimes \vec{z_3} \otimes \vec{z_4}
                                                                       \otimes \vec{z_5} \otimes \vec{z_6}\)
contains all the \(15\) combinations of two \(\vec{x}\)s and four \(\vec{z}\)s.

A sufficient condition for 
an \(N\)-qubit state \(\rho\) to have a local realistic model for 
\(N\)-qubit correlations (of two-settings   Bell experiments) is  that
\[\sum_{x_1, \ldots, x_N = x, y} T^2_{x_1 \ldots x_N} \leq 1\] 
for \emph{any} set of local coordinate systems \cite{ZB}. Here \(x\) and \(y\) denote 
any two arbitrary orthogonal directions, which can be separately defined for each observer.

For the noisy \(G_6\) 
state \(\rho^{G_6}\) given by the eq.(\ref{noisyGN}),
\[\sum_{x_1,  \ldots, x_6 = x, y} T^2_{x_1 \ldots x_6}  = \frac{16p_6^2}{3},\]
for the \emph{fixed} local coordinate system in which the secret sharing is considered. 
 Thus
an explicit local realistic description for the \((x,y)\) correlations 
for the state \(\rho^{G_6}\) exists for \[p_6\leq \sqrt{3/16} \approx 0.433012\] 
whereas (for the correlations in the same sector) the  noisy \(6\)-qubit GHZ state (eq.(\ref{noisyGHZ}))
has no local realistic description for \[q_6 > 1/\sqrt{32} \approx 0.176777.\] The latter follows from the fact
that the \(6\)-qubit GHZ state violates the multiqubit Bell inequalities maximally,
and the amount of violation is \(\sqrt{2^{5}}\). This shows that the \(N\)-qubit correlations 
of the GHZ state are stronger in comparison to that of the \(G\)-state.
Since the scheme of the secret sharing protocol is for \(\sigma_x\) and \(\sigma_y\) measurements in a fixed local 
coordinate system, it is probably enough
to check the sufficiency for local realism in the \((x,y)\) sector. However
we show below that the stronger \(N\)-qubit correlations of the GHZ state as compared to 
the \(G\) state persists even when \emph{any} local coordinate systems are considered.

For \emph{any} set of 
local coordinate systems, 
\[\sum_{x_1, \ldots, x_6 = x, y} T^2_{x_1 \ldots x_6} 
\leq \sum_{x_1, \ldots, x_6 = x, y, z} T^2_{x_1 \ldots x_6} =23,\] for the state \(G_6\).  
So the noisy \(G_6\)-state
 has for sure a local realistic description for \(6\)-qubit correlations, for 
\[p_6 \leq 1/\sqrt{23} \approx 0.208514.\] 
Therefore  the \(6\)-qubit GHZ state is definitely more robust to white noise admixture than 
the state \(G_6\) \emph{when \(6\)-qubit correlations are considered}.

\subsection{Higher order correlations are determined by lower order ones for the \(\left|G_N\right\rangle\) state}

We would now give yet another argument to show that the \(N\)-qubit state \(\left|G_{N}\right\rangle\) has 
weaker \(N\)-qubit correlations than the \(N\)-qubit GHZ state. Note that 
for the GHZ state, 
information regarding the reduced density matrices of lower number of parties is insufficient 
to describe the whole \(N\)-party state. Indeed the state 
\[\frac{1}{2}(\left|0^{\otimes N}\right\rangle \left\langle 0^{\otimes N} \right| + 
\left|1^{\otimes N}\right\rangle \left\langle 1^{\otimes N} \right|),\]
has the same single-party, \(2\)-party, \(\ldots\), \((N - 1)\)-party reduced density matrices as the 
state \(\left|\mbox{GHZ}_{N}\right\rangle\). Consequently, high order correlations
of the GHZ states are not determined by lower order correlations. 
However, as shown in \cite{LPW}, this feature is rare.  For almost
all states, actually the opposite is true.

On the lines 
of \cite{LPW}, we will now show that 
  for the states \(\left|G_N\right\rangle\), 
for \(N \geq 5\), there is no other state (pure or mixed)
whose \((N -1)\)-party reduced density matrices match those of the \(G_N\) state.

For convenience of notation, let us call the parties sharing the \(N\)-qubit \(G_N\) state
as \(1, 2, \ldots, N\). A state whose \((N  - 1)\)-party reduced density matrices agree with those 
of \(G_N\), may be a mixed state. We therefore allow for an environment \(E\). 
Any such state, whose \((N-1)\)-party reduced density matrix of the parties \(1, 2, \ldots, (N-1)\) agree
with the state \(G_N\), 
can be written as 
\begin{equation}
\label{12N-1}\left|\chi\right\rangle = \frac{1}{\sqrt{2}}(\left|v_0\right\rangle_{12 \ldots (N-1)} 
                                                           \left|E_0\right\rangle_{NE}
                                       + \left|v_1\right\rangle_{12 \ldots (N-1)} \left|E_1\right\rangle_{NE}),
\end{equation}
where
\begin{equation}
\label{v}
\begin{array}{rcl}
\left|v_0\right\rangle & = & \frac{1}{\sqrt{N}}(\sum \left|10^{\otimes N -2}\right\rangle
                                              + \left|1^{\otimes N - 1}\right\rangle) \\
\left|v_1\right\rangle & = & \frac{1}{\sqrt{N}}(\sum \left|01^{\otimes N -2}\right\rangle
                                             +  \left|0^{\otimes N - 1}\right\rangle) 
\end{array}
\end{equation}
and \(\left|E_0\right\rangle\) and \(\left|E_1\right\rangle\) are orthonormal.
Note that \(\left|v_0\right\rangle \) and \(\left|v_1\right\rangle\)
are just the states \(\left|\xi\right\rangle\) and \(\left|\overline{\xi}\right\rangle\) extended also to the 
case of even \(N\) (see eq. 
(\ref{+Z})).
Since \(\left|E_0\right\rangle\) and \(\left|E_1\right\rangle\) are orthonormal, they can be written as 
\begin{equation}
\label{E}
\begin{array}{rcl}
\left|E_0\right\rangle & = & \left|0\right\rangle_N \left|e_{00}\right\rangle_E 
                           +   \left|1\right\rangle_N \left|e_{01}\right\rangle_E  \\
\left|E_1\right\rangle & = & \left|0\right\rangle_N \left|e_{10}\right\rangle_E 
                           +   \left|1\right\rangle_N \left|e_{11}\right\rangle_E. 
\end{array}
\end{equation}
Let us now try to match the \((N-1)\)-party reduced density matrix of the parties \(2, 3, \ldots, N\) of 
the state \(G_N\) with that of \(\left|\chi\right\rangle\). Consequently, it should be possible to write 
\(\left|\chi\right\rangle\)
as 
\begin{equation}
\label{23N}\left|\chi\right\rangle = \frac{1}{\sqrt{2}}(\left|v_0\right\rangle_{23 \ldots N} 
                                                           \left|F_0\right\rangle_{1E}
                                       + \left|v_1\right\rangle_{23 \ldots N} \left|F_1\right\rangle_{1E}),
\end{equation}
where the states
\(\left|v_0\right\rangle\) and \(\left|v_1\right\rangle\) are  given by eq. (\ref{v}),
and the orthonormal environment states \(\left|F_0\right\rangle\) and \(\left|F_1\right\rangle\)
are given by
\begin{equation}
\label{F}
\begin{array}{rcl}
\left|F_0\right\rangle & = & \left|0\right\rangle_1 \left|f_{00}\right\rangle_E 
                           +   \left|1\right\rangle_1 \left|f_{01}\right\rangle_E  \\
\left|F_1\right\rangle & = & \left|0\right\rangle_1 \left|f_{10}\right\rangle_E 
                           +   \left|1\right\rangle_1 \left|f_{11}\right\rangle_E. 
\end{array}
\end{equation}

The states on the right hand sides of the equations (\ref{12N-1}) and (\ref{23N}) are actually
the same states. Comparing the different terms, it is not hard to see that for \(N \geq 5\),
\begin{equation}
\label{uni}
\begin{array}{lcl}
\left|e_{00}\right\rangle = \left|e_{11}\right\rangle \\
\left|e_{01}\right\rangle = \left|e_{10}\right\rangle = 0.
\end{array}
\end{equation}
Therefore the 
state \(\left|\chi\right\rangle\), for \(N \geq 5\),  is  
a product state of \(\left|G_N\right\rangle\) with a state of the environment,
if \(\left|\chi\right\rangle\)  has its \((N-1)\)-party 
reduced density matrices equal to the corresponding ones for \(\left|G_N\right\rangle\). 
Thus for \(N \geq 5\), there are no other \(N\)-qubit states whose \((N-1)\)-qubit reduced density matrices 
are equal to those of \(\left|G_N\right\rangle\). Therefore, contrary to the GHZ states, given the \((N-1)\)-qubit 
reduced density matrices, the state \(\left|G_N\right\rangle\) is already completely specified. This again
underlines the fact that the state \(\left|G_N\right\rangle\) does not possess strong \(N\)-qubit correlations.
This also explicitly shows that 
the states \(\left|G_N\right\rangle\), for \(N \geq 5\), are not from the GHZ family.

\section{Summary}

We have given a criterion on security of secret sharing protocols.
The criterion is based on violation in a specific way, of Bell inequalities (as given by items (i) and (ii) in the 
Introduction). 
A criterion for security of secret sharing based on violation of Bell inequalities was provided 
in \cite{GS1}. Let us mention here that we do not simply extend this previously known criterion of Ref. \cite{GS1}. 
The criterion mentioned in Ref. \cite{GS1} (call it C1), is not satisfied by the G-states considered in our paper.
 And yet secret sharing is possible by using the G-states. We have 
put forward an independent criterion. Our criterion (call it C2), 
although again based on violation of Bell inequalities, is drastically different from the
 criterion mentioned in Ref. \cite{GS1}. And we 
find that both the GHZ state (which was shown in Ref. \cite{GS1}
 to satisfy C1 and useful for secret sharing) and the G-state 
(which is shown in this paper to violate C1 and still useful for secret sharing) 
are satisfying our criterion C2.  

We believe that these considerations would be useful in the general discussions 
as well as as for 
applications of quantum cryptography \cite{rmp}.

\begin{acknowledgments}

AS and US acknowledges the University of Gda\'{n}sk,
 Grant No. BW/5400-5-0236-2 and BW/5400-5-0256-3. MZ is supported by 
Professorial Subsidy of Foundation for Polish Science.

\end{acknowledgments}


\begin{thebibliography}{9}
\bibitem{robust}Here one must note that secret sharing by the GHZ state has the 
added feature that if any one of the Bobs are left out, the remaining Bobs are with 
strictly no information. We however do not deal with this extreme case. In our protocol, 
if any of the Bobs are left out, the remaining Bobs cannot obtain the complete information
about Alice's bit.
\end{thebibliography}
\end{document}